\documentclass{aa}
\usepackage{times}
\usepackage{graphicx}
\begin{document}

\title{Sodium abundances in nearby disk stars \thanks{Based on observations
collected at the Germany-Spanish Astronomical Center, Calar Alto, Spain.)}}

\author{J.R. Shi\inst{1,2} \and T. Gehren\inst{2} \and G. Zhao \inst{1,2}}

\offprints{J.R. Shi, \email{sjr@bao.ac.cn}}

\institute {National Astronomical Observatories, Chinese Academy of Sciences,
Beijing 100012, P. R. China \and Institut f\"ur Astronomie und Astrophysik der
Universit\"at M\"unchen, Scheinerstrasse 1, D-81679 M\"unchen, Germany}

\date{Received date  / Accepted date}

\abstract{We present sodium abundances for a sample of nearby stars. All results
have been derived from NLTE statistical equilibrium calculations. The influence
of collisional interactions with electrons and hydrogen atoms is evaluated by
comparison of the solar spectrum with very precise fits to the \ion {Na} {I}
line cores. The NLTE effects are more pronounced in metal-poor stars since the
statistical equilibrium is dominated by collisions of which at least the
electronic component is substantially reduced. The resulting influence on the
determination of sodium abundances is in a direction opposite to that found
previously for Mg and Al. The NLTE corrections are about $-0.1$ in thick-disk
stars with [Fe/H] $\sim-0.6$. Our [Na/Fe] abundance ratios are about solar for
thick- and thin-disk stars. The increase in [Na/Fe] as a function of [Fe/H] for
metal-rich stars found by Edvardsson et al. (\cite{EAG93}) is confirmed. Our
results suggest that sodium yields increase with the metallicity, and quite
large amounts of sodium may be produced by AGB stars. We find that [Na/Fe]
ratios, together with either [Mg/Fe] ratio, kinematic data or stellar
evolutionary ages, make possible the individual discrimination between thin- and
thick-disk membership. }

\maketitle

\section{Introduction}

The $^{23}$Na nucleus contains an extra neutron that makes its synthesis deviate
from that of the $\alpha$-nuclei such as $^{24}$Mg. So the abundance ratios
[Na/Fe] and/or [Na/Mg] in stars are a potential source of information concerning
the history of Galactic nucleosynthesis.

It is expected that sodium is made during carbon and neon burning
in massive stars, and it would therefore be ejected by Type II
supernova (SN II). Type Ia supernova, on the other hand, produce
little sodium (Tsujimoto et al. \cite{TNY95}). However, $^{23}$Na
can be synthesized as a primary or secondary element. In the first
case $^{23}$Na would be produced directly in the carbon burning
process, and the production rate would be independent of the metal
content of the proceeding stellar generation. In the second
process, where $^{23}$Na is produced by an excess of neutrons, the
production rate would depend on that neutron excess which in turn
would be determined by the pre-existing metal abundance. In the
latter case $^{23}$Na would be expected to display an
underabundance compared with Fe or Mg in the most metal-poor
stars. Although significant progress has been made in
nucleosynthesis calculations, theoretical SN yields of sodium have
not yet converged to consistency among models from different
authors (Woosley \& Weaver \cite{WW95}; Limongi et al.
\cite{LSC98}; Umeda et al. \cite{UNN00}). Also some synthesis of
sodium in intermediate-mass stars can occur in the
hydrogen-burning shell through the neon-sodium cycle (Denissenkov
\& Denissenkova \cite{DD90}; Mowlavi \cite{M99}).

The observational picture for this element is complicated. The early studies
report that sodium scales with iron at all metallicities (e.g. the review of
Timmes et al. \cite{TWW95} and Goswami \& Prantzos \cite{GP00}). However, new
studies do not support this classical picture. For metal-poor stars, Pilachowski
et al. (\cite{PSK96}) analyzed a sample of 60 halo subgiants, giants and
horizontal branch stars in the interval $-3 \leq$ [Fe/H] $\leq -1$, and found a
small [Na/Fe] deficiency of $-0.17$ dex in the mean and that bright field halo
giants do not show the excess of sodium found in their globular cluster
counterparts. Baum\"{u}ller et al. (\cite{BBG98}) performed a NLTE analysis of
sodium and reported a clear trend of decreasing [Na/Fe] with decreasing
metallicity. Similarly, Stephens (\cite{S99}) found that [Na/Fe] decreases as
one goes from [Fe/H]$=-1$ to [Fe/H]$=-2$, and thus shows the theoretically
expected behaviour of odd-Z elements. Carretta et al. (\cite{CGS00}) and Gratton
et al. ({\cite{GCC03}) also found that sodium is overdeficient in stars with
[Fe/H] $\leq -1$; this result was confirmed by very recent work of Gehren et al.
(\cite{GLS04}).

For disk stars, Edvardsson et al. ({\cite{EAG93}) used
high-resolution, low-noise spectra to determine sodium abundace in
189 F and G stars. Their LTE analysis suggested that the [Na/Fe]
ratio is solar and varies very little with metallicity, which was
confirmed by Chen et al. (\cite{CNZ00}). Based on 5682/5688 and
6154/6160 lines, Prochaska et al. (\cite{PNC00}) measured sodium
abundance in their thick-disk stars and found that all the stars
show mildly enhanced [Na/Fe] and there is a mild trend with
metallicity (see also Gehren et al. \cite{GLS04}).

Concerning the most metal-rich stars in the galactic disk, the
study by Edvardsson et al. ({\cite{EAG93}) raised a number of new
questions. They found sodium relative to iron versus [Fe/H] shows
an increase for [Fe/H] $>0.0$ (see also Feltzing \& Gustafsson
\cite{FG98}, their Fig. 8). Is the ``upturn" real? Large
star-to-star scatter was encountered. Could this scatter to be
reduced by using better abundance criteria?

It is known that [Fe/H] distributions of the halo, thick- and
thin-disk stars reveal considerable overlap, so it can not allow
unequivocal classification for many stars without reference to
other criteria such as age or kinematics. Fuhrmann (\cite{F98},
\cite{F02}) suggested that the thick-disk stars represent the
principal population for the early Galactic evolution, while the
stellar halo contributes by only minute amounts (see also Ryan \&
Smith \cite{RS03}). Considering the general kinematics, there may
be a significant overlap of all populations (Gehren et al.
\cite{GLS04}). Gratton et al. (\cite{GCC96}, \cite{GCM00},
\cite{GCC03}) have shown that thick-disk stars are more [Mg/Fe]
enhanced than the thin-disk stars at the same [Fe/H] value. This
trend has also been reported by Fuhrmann (\cite{F98}; see also
Prochaska et al. \cite{PNC00} and Reddy et al. \cite{RTL03}).
Finally, Mashonkina et al. (\cite{MGT03}) have given evidence that
[Eu/Mg] ratio allows the individual discrimination of halo stars
from thick-disk stars, and similar behaviour is found for the
[Al/Mg] ratio by Gehren et al. (\cite{GLS04}).

The present work is based on a sample of nearby stars and aims at exploring
their [Na/Fe] abundance ratios applying full spectrum synthesis based on level
populations calculated from the statistical equilibrium equations. In Sect. 2 we
present the observational techniques, while the atmospheric models and stellar
parameters are discussed in Sect. 3. NLTE line formation is discussed in Sect.
4. The discussion is presented in Sect. 5, and the conclusions are found in the
last section.

\section{ Observations}

Our approach towards a representative abundance investigation is
aimed at analyzing a volume-complete sample of the F, G and early
K dwarfs in the disk population, north of declination $\delta =
-15^{\circ}$ and within 25pc of the Sun (see Fuhrmann \cite{F04},
for details). The spectra of our samples were obtained through the
years 1995 to 2000 by Klaus Fuhrmann with the fiber-coupled
Cassegrain \'{e}chelle spectrograph FOCES (Pfeiffer et al.
\cite{PFB98}) mounted at the 2.2m telescope of the Calar Alto
Observatory. Only few of them were exposed to a 1024$^2$ 24$\mu$
CCD with $\lambda / \Delta\lambda \sim 40000$, and their
wavelength ranges are limited to 4000-7000 \AA, whereas the
standard configuration was a 2048$^2$ 15$\mu$ CCD that covered
4000-9000 \AA\ with $\lambda/ \Delta\lambda \sim 60000$. All stars
were observed at least twice with a signal-to-noise ratio of at
least S/N $\sim$ 150 up to S/N $\sim 400$ (see Fuhrmann
\cite{F98}, \cite{F00}, for details).

\section {Atmospheric models and stellar parameters}
\subsection{Model atmospheres}
Our analyses are all based on the same type of atmospheric model, irrespective
of temperature, gravity or metal abundance. We use line-blanketed LTE model
atmospheres, generated as discussed by Fuhrmann et al. (\cite{FP97}). The main
characteristics are:
\begin{itemize}
\item
the iron opacity was calculated with the improved meteoritic value log
$\varepsilon_{\rm Fe} = 7.51$
\item
opacities for metal-poor stars with [Fe/H] $< -0.6$ were calculated using
$\alpha$-element (O, Ne, Mg, Si, S, Ar, Ca and Ti) abundances enhanced by 0.4 dex
\item
the mixing-length parameter $l$/H$_p$ was adopted to be 0.5 (see Fuhrmann et al.
\cite{FAG93})
\end{itemize}

\subsection{Stellar parameters}

For most of the stars we adopt the stellar parameters determined
by Fuhrmann (\cite{F98}, \cite{F00}), where the effective
temperatures are derived from the wings of the Balmer lines;
surface gravities are taken from the strong line wings of the
\ion{Mg}{I}b triplet and compared with values obtained from
Hipparcos parallaxes. Iron abundances are based on \ion{Fe} {II}
lines, and the microturbulence velocities are estimated by
requesting that the iron abundance derived from \ion{Fe} {II}
lines should not depend on equivalent width. The uncertainties for
the temperature, surface gravity, metal abundance and
microturbulence velocities are generally assumed to be $\pm$80 K,
0.1 dex, 0.2 km s$^{-1}$ and 0.07 dex respectively (see Fuhrmann
\cite{F98} and \cite{F00} for details). For Procyon the parameters
were taken from Korn et al. (\cite{KSG03}), where the temperatures
are again from the wings of Balmer lines; surface gravities are
based on the HIPPARCOS parallaxes, and iron abundances are taken
from NLTE calculations.

Following Fuhrmann's work, we employ the macroturbulence parameters $\zeta$ in
the radial-tangential form and adopt values as described in Gray~(\cite{G84}).
The projected rotational velocity is obtained as the residual to the observed
line profiles, after having removed the known instrumental profile obtained from
the Moon spectra.

\section{NLTE line formation}
\subsection{Atomic model}
The model of Na we adopt goes back to the atomic properties documented in
Baum\"{u}ller et al. (\cite{BBG98}), where most of the problems have been
described in detail. In this paper, we extend the number of levels to $n = 15$.
The model is closed with the Na$^+$ ground state. Fine structure splitting is
neglected with the exception of $3p^2P^0_{1/2}$ and $3p^2P^0_{3/2}$. Part of
that model was shown by Baum\"{u}ller et al. (\cite{BBG98}, see their Fig. 2).
Energy levels are taken from Martin \& Zalubas (1981) if available. For levels
with $l \ge 4$ hydrogenic approximations are used. All bound-free radiative
cross-sections have been included from close-coupling calculations of Butler
(\cite{B93}) and Butler et al. (\cite{BMZ93}).

To establish the influence of deviations from LTE on the ionization equilibrium
of sodium, we have chosen the reference stars of Korn et al. (\cite{KSG03}). The
full analysis of the solar spectrum (Kurucz et al. \cite{KFB84}) and those of
the reference stars allows a reasonable choice of the hydrogen collision
enhancement factor resulting in S$_H$ = 0.05 for \ion{Na}, found by iteration
together with all basic stellar paraeters. This number is the same as in the
previous work of Baum\"{u}ller et al. (\cite{BBG98}).

All calculations have been carried out with a revised version of the DETAIL
program (Butler \& Giddings \cite{BG85}) using accelerated lambda iteration (see
Gehren et al. {\cite{GBM01}, \cite{GLS04} for details).

\subsection {Atomic line data}
As our primary choice we take the oscillator strengths from the NIST data base. We note
that the $gf$ values are similar to those adopted by Takada et al. (\cite{TZT03}), the
average difference log $gf$(this work)$-$ log $gf$(Takeda) is $-0.02\pm0.04$. For the
line pair of 6154/6160, our $gf$ values are same as the values adopted by Chen et al.
(\cite{CNZ00}), but slightly higher than the values of Edvardsson et al. (\cite{EAG93})
and Feltzing \& Gustafsson (\cite{FG98}), the difference is about 0.04. However, in our
present analysis the absolute value of the oscillator strengths is unimportant because
the abundances are evaluated in a fully differential way with respect to the sun.

Collisional broadening through van der Waals interaction with hydrogen atoms is
important for strong Na D lines. As already pointed out by Gehren et al.
(\cite{GBM01}, \cite{GLS04}) the resulting values of the van der Waals damping
constants are mostly near those calculated according to Anstee \& O'Mara's
(\cite{AO91}, \cite{AO95}) tables. Fitting the solar profiles with a common Na
abundance then shows that our damping constants for the resonance lines are
roughly a factor of 2 below those calculated. The adopted line data ($gf$ values
and van der Waals damping constants) are given in Table 1.

\begin{table}
\caption[1]{Atomic data of Na lines. $\log gf$ values and damping constants have
been determined from solar spectrum fits}
\centering{\begin{tabular}{rr@{ $-$ }lrr}
\hline\hline\noalign{\smallskip}
$\lambda$~ [\AA] & \multicolumn{2}{c}{Transition} & $\log gf$ & $\log C_6$\\
\hline\noalign{\smallskip}
 5682.642 &  $3p\,^2{\rm P}^{0}_{1/2}$ & $4d\,^2{\rm D}_{3/2}$ & -0.72  & -30.05  \\
 5688.214 &  $3p\,^2{\rm P}^{0}_{3/2}$ & $4d\,^2{\rm D}_{5/2}$ & -0.47  & -30.05  \\
 5889.959 &  $3s\,^2{\rm S}_{1/2}$ & $3p\,^2{\rm P}^{0}_{1/2}$ & +0.11  & -31.60  \\
 5895.932 &  $3s\,^2{\rm S}_{1/2}$ & $3p\,^2{\rm P}^{0}_{3/2}$ & -0.19  & -31.60  \\
 6154.228 &  $3p\,^2{\rm P}^{0}_{1/2}$ & $5s\,^2{\rm S}_{1/2}$ & -1.57  & -29.78  \\
 6160.751 &  $3p\,^2{\rm P}^{0}_{3/2}$ & $5s\,^2{\rm S}_{1/2}$ & -1.28  & -29.78  \\
 8183.260 &  $3p\,^2{\rm P}^{0}_{1/2}$ & $3d\,^2{\rm D}_{3/2}$ &  0.28  & -30.73  \\
 8194.800 &  $3p\,^2{\rm P}^{0}_{3/2}$ & $3d\,^2{\rm D}_{5/2}$ &  0.49  & -30.73  \\
\noalign{\smallskip}
\hline
\end{tabular}}
\end{table}
\subsection{Sodium abundances}
The abundance determinations for our program stars are made using
spectral synthesis. The synthetic spectra are convolved with
macroturbulence, rotational and instrumental broadening profiles,
in order to match the observed spectral lines. Based on 5682/5688,
5890/5896 and 6154/6160 line pairs, our abundance results are
obtained from profile fits. However, for Na D lines, as pointed
out by Baum\"{u}ller et al. (\cite{BBG98}), even in the solar
atmosphere the inner cores of the lines are affected by deviations
from LTE. We note that even for some metal-poor stars, only the
line wings can be fitted in LTE. So the sodium abundances are
obtained until a best fit to the observed line wings is provided
for the LTE analyses. Considering the NLTE abundances, out results
do not show large abundance differences between different lines
(however, see Takeda et al. \cite{TZT03}). The final abundance
scatter of single line is between 0.01 and 0.06  with
$<\sigma$[Na/Fe]$>=0.020 \pm 0.007$ for all stars. The derived
abundances are presented in Table 2. Complete results for
individual lines of all stars can be obtained from the author on
request.

\begin{table}
\caption[2]{Atmospheric parameters and sodium abundances of the
program stars. Rotational velocities and macroturbulence data are
the same as in Fuhrmann (\cite{F98}, \cite{F00})}
\begin{tabular}{rrrrrrr}
\hline\hline\noalign{\smallskip}
HD & $T_{\rm eff}$ & $\log g$ & [Fe/H] & $\xi$ & [Na/Fe] & [Na/Fe]\\
   &               &          &        &       &     LTE &    NLTE\\
\noalign{\smallskip}\hline\noalign{\smallskip}
    400 &      6149 &     4.06 &  -0.25 &  1.31 &    0.11 &    0.03\\
   3079 &      6050 &     4.17 &  -0.14 &  1.42 &    0.09 &    0.02\\
   4614 &      5939 &     4.33 &  -0.30 &  1.06 &    0.09 &    0.04\\
   5015 &      6045 &     3.90 &  -0.02 &  1.39 &    0.12 &    0.04\\
   6582 &      5387 &     4.45 &  -0.83 &  0.89 &    0.14 &    0.07\\
\noalign{\smallskip}
   6920 &      5838 &     3.48 &  -0.05 &  1.35 &    0.08 &    0.02\\
   9407 &      5663 &     4.42 &   0.03 &  0.87 &    0.09 &    0.05\\
   9826 &      6107 &     4.01 &   0.09 &  1.40 &    0.13 &    0.04\\
  10697 &      5614 &     3.96 &   0.10 &  1.04 &    0.11 &    0.04\\
  16141 &      5737 &     3.92 &   0.02 &  1.24 &    0.14 &    0.06\\
\noalign{\smallskip} \noalign{\smallskip}
  16895 &      6248 &     4.20 &  -0.01 &  1.42 &    0.08 &    0.00\\
  18757 &      5714 &     4.34 &  -0.28 &  0.95 &    0.13 &    0.06\\
  18803 &      5657 &     4.39 &   0.14 &  0.87 &    0.12 &    0.05\\
  19373 &      5966 &     4.15 &   0.03 &  1.23 &    0.24 &    0.16\\
  19994 &      6087 &     3.97 &   0.14 &  1.19 &    0.20 &    0.12\\
\noalign{\smallskip}
  20619 &      5706 &     4.48 &  -0.20 &  0.90 &    0.02 &   -0.04\\
  22879 &      5867 &     4.26 &  -0.84 &  1.21 &    0.06 &    0.00\\
  25457 &      6246 &     4.32 &   0.06 &  1.25 &    0.10 &    0.03\\
  30649 &      5816 &     4.28 &  -0.47 &  1.18 &    0.06 &    0.00\\
  30743 &      6298 &     4.03 &  -0.45 &  1.64 &    0.15 &    0.04\\
\noalign{\smallskip} \noalign{\smallskip}
  34411 &      5875 &     4.18 &   0.03 &  1.08 &    0.13 &    0.07\\
  37124 &      5610 &     4.44 &  -0.44 &  0.89 &    0.17 &    0.11\\
  43042 &      6444 &     4.23 &   0.04 &  1.52 &    0.01 &   -0.06\\
  43386 &      6480 &     4.27 &  -0.06 &  1.56 &    0.26 &    0.16\\
  52711 &      5887 &     4.31 &  -0.16 &  1.04 &    0.10 &    0.04\\
\noalign{\smallskip}
  58855 &      6309 &     4.16 &  -0.32 &  1.38 &    0.14 &    0.04\\
  61421 &      6600 &     3.96 &  -0.03 &  1.83 &    0.19 &    0.11\\
  62301 &      5938 &     4.06 &  -0.69 &  1.28 &    0.11 &    0.04\\
  67228 &      5847 &     3.93 &   0.12 &  1.18 &    0.22 &    0.14\\
  69611 &      5821 &     4.18 &  -0.60 &  1.22 &    0.14 &    0.07\\
\noalign{\smallskip} \noalign{\smallskip}
  75732 &      5336 &     4.47 &   0.40 &  0.76 &    0.28 &    0.25\\
  90508 &      5802 &     4.35 &  -0.33 &  1.02 &    0.03 &   -0.02\\
  95128 &      5892 &     4.27 &   0.00 &  1.01 &    0.12 &    0.06\\
 102158 &      5758 &     4.24 &  -0.46 &  1.13 &    0.14 &    0.08\\
 102870 &      6085 &     4.04 &   0.14 &  1.38 &    0.08 &    0.00\\
\noalign{\smallskip}
 109358 &      5863 &     4.36 &  -0.21 &  1.12 &    0.03 &   -0.02\\
 114710 &      6006 &     4.30 &  -0.03 &  1.12 &    0.08 &    0.03\\
 114762 &      5934 &     4.11 &  -0.71 &  1.14 &    0.13 &    0.07\\
 117176 &      5483 &     3.82 &  -0.09 &  1.02 &    0.04 &   -0.03\\
 120136 &      6360 &     4.17 &   0.27 &  1.56 &    0.07 &    0.00\\
\noalign{\smallskip} \noalign{\smallskip}
 121370 &      6023 &     3.76 &   0.28 &  1.40 &    0.35 &    0.25\\
 121560 &      6144 &     4.27 &  -0.43 &  1.26 &    0.14 &    0.07\\
 122742 &      5537 &     4.43 &  -0.01 &  0.78 &    0.12 &    0.07\\
 124850 &      6112 &     3.78 &  -0.06 &  1.48 &    0.10 &   -0.03\\
 126053 &      5691 &     4.45 &  -0.35 &  0.96 &    0.03 &   -0.02\\
\noalign{\smallskip}
 127334 &      5691 &     4.22 &   0.21 &  0.97 &    0.18 &    0.12\\
 130322 &      5394 &     4.55 &   0.04 &  0.79 &    0.05 &    0.01\\
 130819 &      6598 &     4.18 &  -0.10 &  1.76 &    0.06 &   -0.02\\
 132254 &      6220 &     4.15 &  -0.03 &  1.46 &    0.10 &    0.03\\
        &           &          &        &       &         &        \\
\noalign{\smallskip} \hline
\end{tabular}
\end{table}
\begin{table}
{\bf Table 2} (continued)\\

\begin{tabular}{rrrrrrr}
\hline\hline\noalign{\smallskip}
 HD & $T_{\rm eff}$ & $\log g$ & [Fe/H] & $\xi$ & [Na/Fe] & [Na/Fe]\\
    &               &          &        &       &     LTE &    NLTE\\
\noalign{\smallskip}\hline \noalign{\smallskip}
\noalign{\smallskip}
 134987 &      5740 &     4.25 &   0.25 &  1.02 &    0.24 &    0.17\\
 136064 &      6116 &     3.86 &  -0.05 &  1.41 &    0.18 &    0.11\\
 136202 &      6104 &     3.80 &  -0.09 &  1.45 &    0.21 &    0.12\\
 137052 &      6283 &     3.94 &  -0.03 &  1.89 &   -0.02 &   -0.09\\
 141004 &      5864 &     4.09 &  -0.03 &  1.05 &    0.08 &    0.01\\
\noalign{\smallskip}
 142373 &      5841 &     3.84 &  -0.57 &  1.24 &    0.13 &    0.05\\
 142860 &      6254 &     4.02 &  -0.19 &  1.36 &    0.10 &    0.02\\
 143761 &      5821 &     4.12 &  -0.24 &  1.10 &    0.05 &   -0.02\\
 145675 &      5334 &     4.51 &   0.45 &  0.81 &    0.23 &    0.18\\
 146233 &      5786 &     4.36 &   0.01 &  0.98 &    0.09 &    0.03\\
\noalign{\smallskip} \noalign{\smallskip}
 150680 &      5814 &     3.72 &  -0.03 &  1.38 &    0.25 &    0.16\\
 153597 &      6275 &     4.36 &  -0.11 &  1.48 &   -0.01 &   -0.07\\
 154345 &      5507 &     4.57 &  -0.03 &  0.70 &    0.00 &   -0.04\\
 157214 &      5735 &     4.24 &  -0.34 &  1.00 &    0.12 &    0.06\\
 157347 &      5643 &     4.31 &  -0.01 &  0.90 &    0.08 &    0.02\\
\noalign{\smallskip}
 159222 &      5845 &     4.28 &   0.09 &  1.05 &    0.15 &    0.09\\
 161797 &      5596 &     3.93 &   0.23 &  1.17 &    0.23 &    0.16\\
 162003 &      6423 &     4.06 &  -0.07 &  1.67 &    0.17 &    0.05\\
 162004 &      6203 &     4.24 &  -0.06 &  1.29 &    0.11 &    0.04\\
 163989 &      6116 &     3.62 &  -0.14 &  1.61 &    0.13 &    0.03\\
\noalign{\smallskip} \noalign{\smallskip}
 165401 &      5811 &     4.41 &  -0.39 &  1.10 &    0.10 &    0.04\\
 165908 &      5994 &     4.01 &  -0.61 &  1.25 &    0.17 &    0.09\\
 168009 &      5785 &     4.23 &  -0.03 &  1.03 &    0.13 &    0.06\\
 168151 &      6357 &     4.02 &  -0.33 &  1.68 &    0.05 &   -0.02\\
 173667 &      6305 &     3.99 &   0.03 &  1.60 &    0.09 &   -0.04\\
\noalign{\smallskip}
 176377 &      5860 &     4.43 &  -0.27 &  0.91 &    0.04 &    0.00\\
 178428 &      5673 &     4.23 &   0.12 &  0.96 &    0.10 &    0.04\\
 182572 &      5610 &     4.19 &   0.37 &  1.01 &    0.19 &    0.12\\
 184499 &      5828 &     4.13 &  -0.51 &  1.17 &    0.14 &    0.05\\
 186408 &      5805 &     4.26 &   0.06 &  1.03 &    0.14 &    0.08\\
\noalign{\smallskip} \noalign{\smallskip}
 186427 &      5766 &     4.29 &   0.05 &  0.89 &    0.14 &    0.08\\
 187013 &      6312 &     4.09 &  -0.05 &  1.46 &    0.02 &   -0.08\\
 187691 &      6088 &     4.07 &   0.07 &  1.35 &    0.13 &    0.04\\
 187923 &      5713 &     3.97 &  -0.20 &  1.19 &    0.16 &    0.08\\
 190360 &      5588 &     4.27 &   0.24 &  0.98 &    0.15 &    0.09\\
\noalign{\smallskip}
 190406 &      5937 &     4.35 &  -0.01 &  1.10 &    0.10 &    0.05\\
 204155 &      5829 &     4.12 &  -0.63 &  1.18 &    0.12 &    0.03\\
 207978 &      6313 &     3.94 &  -0.52 &  1.57 &    0.20 &    0.10\\
 209458 &      6082 &     4.33 &  -0.06 &  1.15 &    0.05 &    0.00\\
 210277 &      5541 &     4.42 &   0.26 &  0.73 &    0.07 &    0.02\\
\noalign{\smallskip}
 \noalign{\smallskip}
 210855 &      6176 &     3.81 &   0.21 &  1.61 &    0.11 &    0.02\\
 215648 &      6110 &     3.85 &  -0.37 &  1.69 &    0.21 &    0.10\\
 216385 &      6212 &     3.85 &  -0.26 &  1.54 &    0.15 &    0.05\\
 217014 &      5793 &     4.33 &   0.20 &  0.95 &    0.19 &    0.12\\
 218470 &      6407 &     4.07 &  -0.12 &  1.85 &    0.05 &   -0.08\\
\noalign{\smallskip}
 219623 &      6140 &     4.23 &   0.01 &  1.30 &    0.11 &    0.05\\
 221830 &      5749 &     4.19 &  -0.36 &  1.14 &    0.02 &   -0.04\\
 222368 &      6157 &     3.95 &  -0.19 &  1.51 &    0.15 &    0.06\\
  G78-1 &      5895 &     4.11 &  -0.88 &  1.14 &    0.17 &    0.07\\
        &           &          &                &         &        \\
        &           &          &                &         &        \\
        &           &          &                &         &        \\

\noalign{\smallskip} \hline
\end{tabular}
\end{table}

\subsection{Comparison with other works}
Sodium abundances have been determined by several groups including
both LTE and NLTE analyses. In Fig.1 we compare the [Na/Fe] values
(NLTE) determined in this paper with those from the literature.
The agreement is generally good. However, a closer inspection
reveals some systematic differences. In the remaining part of this
section we will briefly discuss some possible reasons for these
differences.

\vskip 0.2cm
 \noindent{\underline{Baum\"{u}ller et al. (\cite{BBG98})}}
\vskip 0.1cm

Baum\"{u}ller et al. (\cite{BBG98}) performed a full NLTE line formation for a
sample of metal-poor stars. We have four stars in common with theirs. Our NLTE
results are very much in agreement with theirs, as shown in Fig. 1. The average
differences between our [Na/Fe] and theirs are 0.02 $\pm$ 0.10. We note that the
considerable scatter is mostly due to the different stellar parameters adopted.

\begin{figure}
\resizebox{\hsize}{7.0cm}{\includegraphics[width=9.3cm]{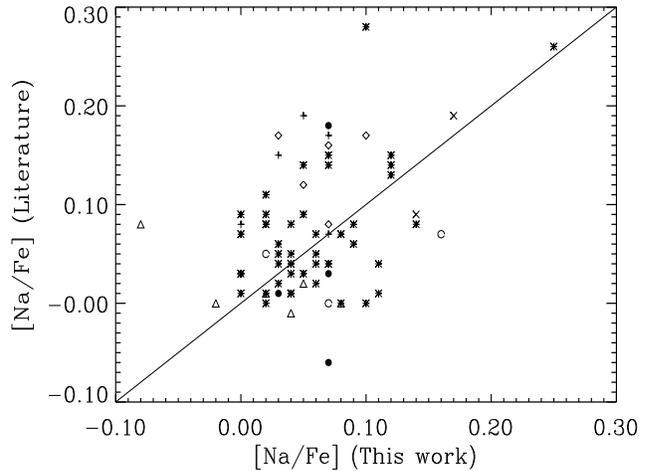}}
\caption[short title]{Comparison of derived [Na/Fe] for stars in
common with other studies. Filled circles ($\bullet$) are from
Baum\"{u}ller et al. (\cite{BBG98}); crosses ($\times$) are from
Feltzing \& Gustafsson (\cite{FG98}); asterisks ($\ast$) are from
Carretta et al. (\cite{CGS00}); pluses ($+$) are from Fulbright
(\cite{Fu00}); diamonds ($\diamond$) are from Gratton et al.
(\cite{GCC03}); open circles ($\circ$) are from Takeda et al.
(\cite{TZT03}); triangles ($\triangle$) are from Reddy et al.
(\cite{RTL03}).}
\vspace*{-0.3cm}
\end{figure}

\vskip 0.2cm
\noindent{\underline{Feltzing \& Gustafsson (\cite{FG98})}}
\vskip 0.1cm

In this study of 47 G and K metal-rich stars, they confirmed the
increase of [Na/Fe] as a function of [Fe/H] found previously by
Edvardsson et al. (\cite{EAG93}). Their results were determined
from 6154/6160 lines, the $gf$ values adopted in this study are
slightly lower than ours, the differences are $-0.02/-0.05$,
respectively. The sodium abundance determined by this study is in
good agreement with ours. For three stars in common, the average
difference is $0.01\pm0.04$.

\vskip 0.2cm
\noindent {\underline{Carretta et al. (\cite{CGS00})}}
\vskip 0.1cm

This work reanalysed the data of Edvardsson et al. (\cite{EAG93}) including a
NLTE correction. They found no large differences from the LTE results of
Edvardsson et al., which is shown in their Fig. 9. The $gf$ values of the
6154/6160 lines adopted by Edvardsson et al. (\cite{EAG93}) are slightly lower
than ours, the differences are $-0.04$ and $-0.03$, respectively. The results of
Carretta et al. (\cite{CGS00}) are very much in agreement with ours. For the 46
common stars from Edvardsson et al. (\cite{EAG93}), the average difference
between ours and theirs is $-0.01\pm 0.05$.

\vskip 0.2cm
\noindent{\underline{Fulbright (\cite{Fu00}, \cite{Fu02})}}
\vskip 0.1cm

This analysis deals with a large number of metal-poor stars, of
which five stars are in common with our sample. As pointed out by
Gehren et al. (\cite{GLS04}), there are large differences in
stellar parameters; on average, the typical difference in
temperatures for the individual star is about 150 K, and there are
a few stars for which differences may be much larger. Fulbright's
abundances are slightly larger than ours (see Fig. 1). For the
five stars in common with our list we obtain $\overline{\Delta
\rm[Na/Fe]}$=$-0.09\pm0.05$. We find that the largest difference
($\sim$0.4 dex) comes from warm metal-poor stars such as
HD\,284248. We note that considerable discrepancies remain even
after correcting for the difference in effective temperatures. As
these stars are metal-poor, they may suffer from relatively large
NLTE effects (see Sect 5.1).

\vskip 0.2cm
\noindent{\underline{Gratton et al.(\cite{GCC03})}}
\vskip 0.1cm

The authors considered NLTE corrections and reported the sodium abundances for
150 field subdwarfs and subgiants. Our results are mostly in agreement with
theirs. For the six stars in common, the average difference is $0.07\pm0.05$.

\vskip 0.2cm
\noindent{\underline{Takeda et al. (\cite{TZT03})}}
\vskip 0.1cm

Very recently, Takeda et al. (\cite{TZT03}) performed a NLTE analysis of sodium
abundances for disk and halo stars. They reanalysed the data from Chen et al.
(\cite{CNZ00}). Their results vary from line to line (see their Tables 4 and 5).
The results for the 6154/6160 lines show no large differences from those of Chen
et al. (see their Fig. 5c). Our [Na/Fe] values are in agreement with theirs for
the lines of 6154/6160 \AA\ for the three stars in common; the average
difference is $0.04\pm0.06$. As already pointed out by Takeda et al., this line
pair is not very sensitive to the NLTE effects.

\vskip 0.2cm
\noindent{\underline{Reddy et al. (\cite{RTL03})}}
\vskip 0.1cm

This sample consists of 187 F and G dwarfs. The sodium abundances
were based on 6154/6160 lines, the $gf$ values adopted in their
work are nearly same as those we adopted. A very close star by
star agreement between the [Na/Fe] ratios derived here and those
measured by Reddy et al., no difference exceeds 0.20 dex. For the
six stars in common, we obtain $\overline{\Delta
\rm[Na/Fe]}$=0.00$\pm 0.08$. The large scatter is due to
HD\,218470 (0.16 dex). This is a high $v \sin i$ (13.1 km
s$^{-1}$), high temperature ($T_{\rm eff}=6407$) and low gravity
(log $g=4.02$) star. For this star their temperature is about 70 K
higher than ours.

\section {Discussion}
\subsection{NLTE effects}

The abundance analyses of sodium clearly show the NLTE effect.
There is a tendency that the NLTE effect is large for warm
metal-poor subgiant stars, as would be expected. Baum\"{u}ller et
al. (\cite{BBG98}) show in their Fig. 11 that the NLTE line cores
for the Na D lines are much deeper than in LTE in metal-deficient
stars. In LTE abundance analyses this is compensated by simply
increasing the sodium abundance until the observed equivalent
width is reproduced. Such results are displayed in Table 3 of
Baum\"{u}ller et al. (see also their Table 2). The results
confirm, starting from the solar abundance towards the lower
metallicity, that the NLTE abundance effect resulting from the Na
D lines increases reaching a maximum near [Fe/H]$=-2$. LTE
abundances can be significantly different from their NLTE
counterparts, with differences reaching more than 0.6 dex in
extreme cases. In Table 3 it is evident that the NLTE effects are
systematically stronger for the hotter models, which is in
agreement with the statistical equilibrium of both aluminium and
magnesium (Baum\"{u}ller et al. \cite{BG97}; Zhao \& Gehren
\cite{ZG00}). As expected, the strongest departures from LTE are
found for models with high temperature and low metallicity. The
reduction of surface gravity results in a decreased efficiency of
collisions with electrons and hydrogen atoms, which again leads to
stronger NLTE effects. In Fig. 2, the differences between LTE and
NLTE analyses for our program stars are plotted as a function of
metal abundance, temperature and luminosity.

It should be noted that the NLTE effects differ from line to line, reflecting
their individual properties. For example, the 8183/8194 lines show relatively
large NLTE effects compared to the other lines at similar line-strength. From
Table 3, we can see the importance of the NLTE effect on different lines in
sodium abundance determination:

\begin{figure}
\resizebox{\hsize}{6.6cm}{\includegraphics[width=9.3cm]{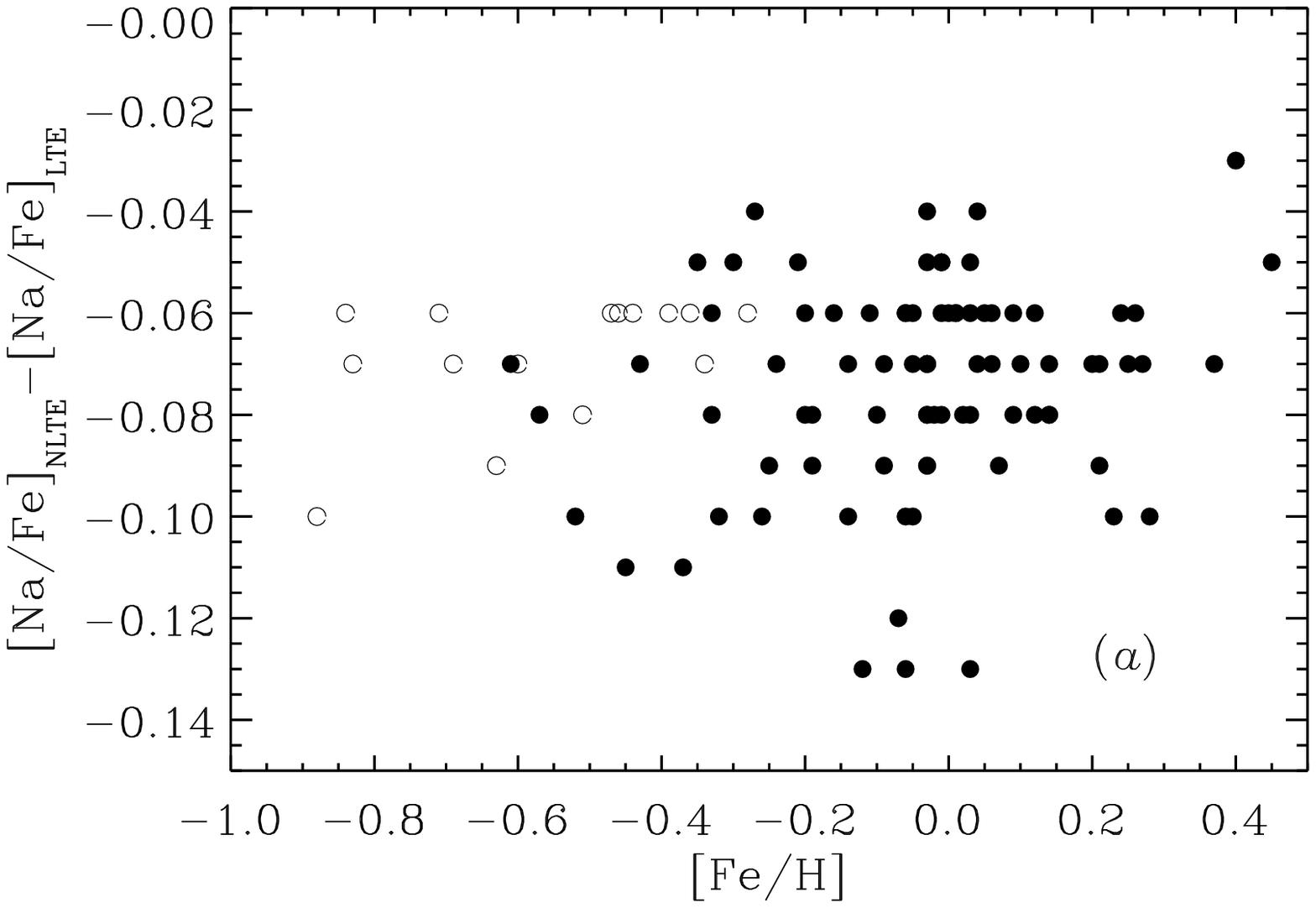}}
\resizebox{\hsize}{6.6cm}{\includegraphics[width=9.3cm]{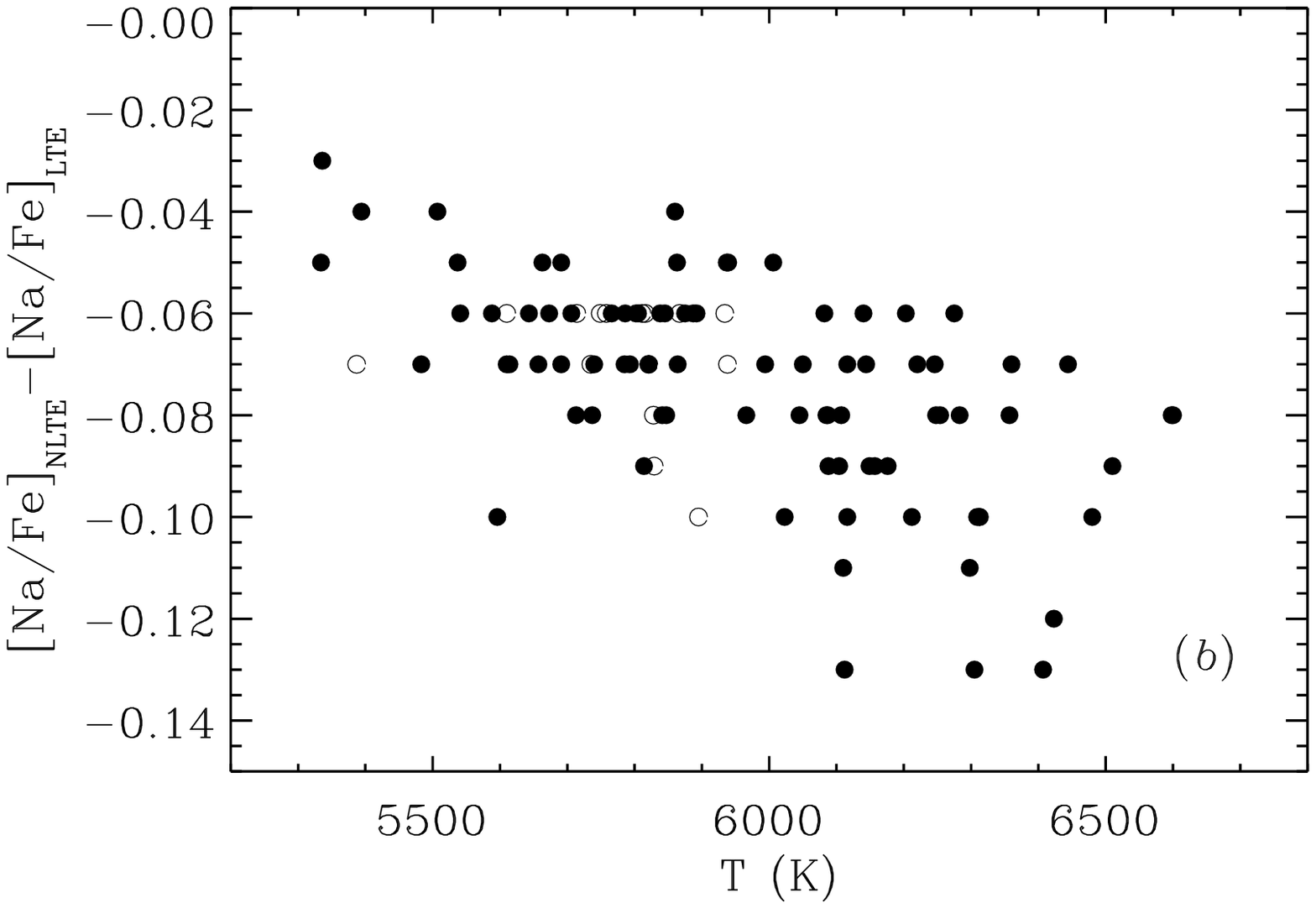}}
\resizebox{\hsize}{6.6cm}{\includegraphics[width=9.3cm]{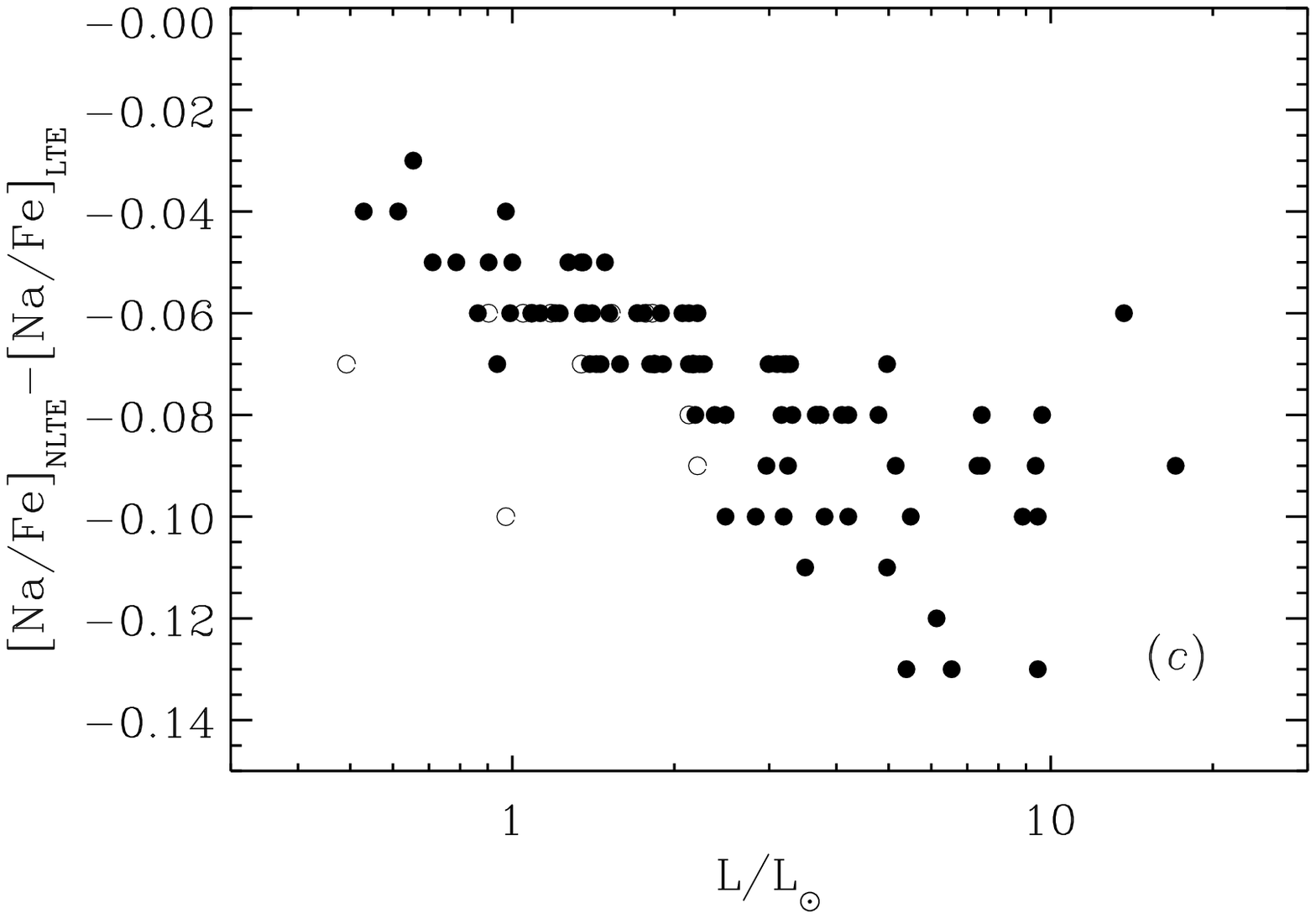}}

\caption[short title]{Difference of [Na/Fe] abundance ratios
calculated under NLTE and LTE assumptions as a function of metal
abundance (a), temperature (b) and luminosity (c). Filled circles
refer to the thin-disk stars, open circles refer thick-disk
stars.}
\end{figure}

\begin{itemize}
\item
the weak 6154/6160 lines show the smallest NLTE abundance effects
($\leq 0.1$ dex) for the disk stars, so they are the best abundance indicators
for LTE analyses of moderately metal-poor stars.
\item
the NLTE corrections for 5682/5688 lines are relatively small ($\sim0.1$ dex)
for dwarfs; but they increase for subgiants ($>0.1$ dex). Therefore these lines
should no longer be analyzed with LTE.
\item
for the 5890/5896 lines, which are important for determining the sodium abundances
of very metal-deficient stars, it is necessary to take account of the NLTE effects
except for very cool and metal-rich dwarfs. Especially for warm metal-poor giants,
the NLTE correction can reach 0.6 dex (see Table 3 for details).
\item
the 8183/8194 lines are also important for determining the sodium abundances of
very metal-deficient stars because of their strength. We suggest that, even for
cool metal-rich stars, the NLTE effects should be taken into account.
\end{itemize}

We confirm that departures from LTE of the Na level populations appear to be
larger for warm and giant stars, which was also found by Baum\"{u}ller et al.
(\cite{BBG98}) and Takeda et al. (\cite{TZT03}).

A warning is appropriate that Table 3 should not be used for LTE
abundance corrections if precise Na abundances are aimed at. The
data may vary considerably whenever the microturbulence velocity
deviates from 1 km\,s$^{-1}$, and any simple correction may result
in an increased scatter. The reference to [Na/Fe] = 0 may also be
a bad approximation, particularly in metal-poor stars.

\begin{table*}
\caption[3]{Abundance differences between NLTE and LTE obtained by fitting
calculated LTE \emph{equivalent widths} of \ion{Na}{I} lines using the same
parameters for various stellar model atmospheres based on the line-blanketed
model grid. Results refer to $\Delta$[Na/Fe]$=$ [Na/Fe]$_{\rm NLTE}-$
[Na/Fe]$_{\rm LTE}$. No entries are given for extremely weak lines. The
calculations all refer to $\xi = 1$ km s$^{-1}$ and [Na/Fe] = 0} \vspace{0.1cm}
\centering{
\begin{tabular}{rrrrrrrrrrrc}
\hline\hline\noalign{\smallskip}
$T_{\rm eff}$  & $\log g$ & [Fe/H]
                           &   5682  &  5688 &  5890 &  5896 &  6154 &  6160 &  8183 &  8194 &
                                                                               $\overline{\Delta{\rm [Na/Fe]}}$\\
\noalign{\smallskip}\hline\noalign{\smallskip}
 5200  &   4.50  &    0.0  &  -0.08  & -0.07 & -0.04 & -0.04 & -0.04 & -0.05 & -0.15 & -0.15 & -0.08 $\pm$ 0.05\\
 5200  &   4.50  &   -1.0  &  -0.06  & -0.06 & -0.07 & -0.08 & -0.04 & -0.04 & -0.24 & -0.24 & -0.10 $\pm$ 0.07\\
 5200  &   4.50  &   -2.0  &         &       & -0.14 & -0.17 &       &       &       &       & -0.15 $\pm$ 0.02\\
 5200  &   4.50  &   -3.0  &         &       & -0.17 & -0.17 &       &       &       &       & -0.17 $\pm$ 0.00\\
\noalign{\smallskip}
 5200  &   3.50  &    0.0  &  -0.14  & -0.14 & -0.04 & -0.06 & -0.06 & -0.06 & -0.31 & -0.31 & -0.14 $\pm$ 0.09\\
 5200  &   3.50  &   -1.0  &         &       & -0.14 & -0.16 &       &       & -0.42 & -0.46 & -0.29 $\pm$ 0.15\\
 5200  &   3.50  &   -2.0  &         &       & -0.31 & -0.34 &       &       &       &       & -0.32 $\pm$ 0.02\\
 5200  &   3.50  &   -3.0  &         &       & -0.36 & -0.28 &       &       &       &       & -0.32 $\pm$ 0.04\\
\noalign{\smallskip}
 5780  &   4.50  &    0.0  &  -0.08  & -0.08 & -0.05 & -0.06 & -0.04 & -0.05 & -0.20 & -0.21 & -0.10 $\pm$ 0.05\\
 5780  &   4.50  &   -1.0  &  -0.06  & -0.07 & -0.16 & -0.18 & -0.05 & -0.05 & -0.27 & -0.29 & -0.14 $\pm$ 0.08\\
 5780  &   4.50  &   -2.0  &  -0.05  & -0.05 & -0.34 & -0.32 & -0.05 & -0.05 & -0.10 & -0.12 & -0.13 $\pm$ 0.10\\
 5780  &   4.50  &   -3.0  &         &       & -0.23 & -0.16 &       &       &       &       & -0.19 $\pm$ 0.04\\
\noalign{\smallskip}
 5780  &   3.50  &    0.0  &  -0.17  & -0.18 & -0.10 & -0.10 & -0.06 & -0.08 & -0.43 & -0.43 & -0.19 $\pm$ 0.12\\
 5780  &   3.50  &   -1.0  &  -0.11  & -0.11 & -0.33 & -0.35 &       &       & -0.39 & -0.43 & -0.29 $\pm$ 0.12\\
 5780  &   3.50  &   -2.0  &         &       & -0.56 & -0.52 &       &       &       &       & -0.54 $\pm$ 0.02\\
 5780  &   3.50  &   -3.0  &         &       & -0.27 & -0.17 &       &       &       &       & -0.22 $\pm$ 0.05\\
\noalign{\smallskip}
 6500  &   4.50  &    0.0  &  -0.07  & -0.09 & -0.11 & -0.15 & -0.03 & -0.04 & -0.28 & -0.30 & -0.13 $\pm$ 0.08\\
 6500  &   4.50  &   -1.0  &         &       & -0.38 & -0.41 &       &       & -0.25 & -0.30 & -0.33 $\pm$ 0.06\\
 6500  &   4.50  &   -2.0  &         &       & -0.45 & -0.38 &       &       &       &       & -0.41 $\pm$ 0.04\\
 6500  &   4.50  &   -3.0  &         &       & -0.13 & -0.10 &       &       &       &       & -0.11 $\pm$ 0.02\\
\noalign{\smallskip}
 6500  &   3.50  &    0.0  &  -0.12  & -0.13 & -0.22 & -0.31 & -0.05 & -0.06 & -0.47 & -0.52 & -0.23 $\pm$ 0.15\\
 6500  &   3.50  &   -1.0  &         &       & -0.64 & -0.61 &       &       & -0.32 & -0.40 & -0.49 $\pm$ 0.02\\
 6500  &   3.50  &   -2.0  &         &       & -0.54 & -0.43 &       &       &       &       & -0.49 $\pm$ 0.06\\
 6500  &   3.50  &   -3.0  &         &       & -0.13 & -0.09 &       &       &       &       & -0.11 $\pm$ 0.02\\
\noalign{\smallskip}\hline
\end{tabular}}
\end{table*}

\subsection{Sodium abundance and nucleosynthesis in the Galaxy}

The variation of [Na/Fe] with the stellar metal abundance [Fe/H]
contains information about the chemical evolution of the Galaxy.
Fig. 3 displays the run of [Na/Fe] ratio (calculated in NLTE) with
the overall metal abundance for all stars considered in this
paper. One important feature that can be seen from Fig. 3 is that
the bulk of the sodium abundances scale approximately as iron in
thin-disk stars, with a small overabundance for the thick-disk
stars. The ``upturn" in [Na/Fe] versus [Fe/H] for metal-rich disk
stars, as observed by Edvardsson et al. (\cite{EAG93}), is also
reproduced \footnote{We have 47 stars common with the work of
Edvardsson et al. (\cite{EAG93}). We note that more metal-rich
stars are included in our sample. The upturn in [Na/Fe] vs.
[Fe/H], as observed by Edvardsson et al. is reproduced and
reinforced. We confirm that the upturn is real (also see Feltzing
\& Gustafsson \cite{FG98}).}.

Using magnesium instead of iron as the reference can remove Type
Ia SNe from consideration, because nearly all the sodium and
magnesium are produced in massive stars (Timmes et al.
\cite{TWW95}). Calculations of nucleosynthsis for SNe Ia show that
they should produce only very little amounts of sodium (Nomoto et
al. \cite{NIN97}). The odd-even effects will be enhanced using
magnesium as the basis of comparison. The overall behaviour of
[Na/Mg] ratios versus [Mg/H] is therefore shown in Fig. 4 where
the Mg abundances are taken from Fuhrmann (\cite{F98},
\cite{F00}). We confirm Arnett's (\cite{A71}) prediction that
[Na/Mg] ratios decline with decreasing metallicity. This suggests
that sodium abundances may be more closely coupled to chemical
evolution history, with a gradual enrichment of [Na/Mg] from thick
to thin disk.

\begin{figure}
\resizebox{\hsize}{6.6cm}{\includegraphics[width=9.3cm]{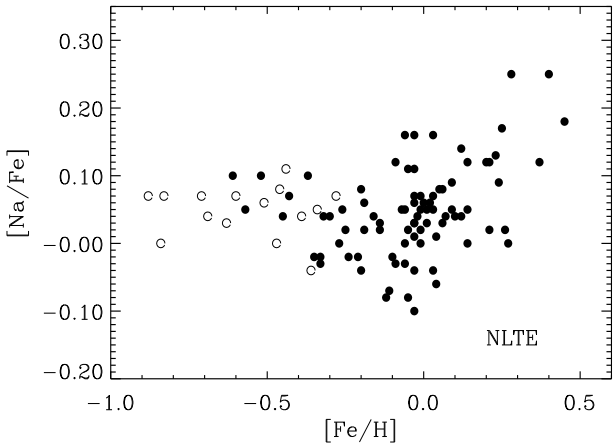}}
\caption[short title]{Abundance ratios [Na/Fe] for NLTE analyses.
The meaning of the symbols is the same as in Fig. 2}
\vspace*{-0.3cm}
\end{figure}

\begin{figure}
\resizebox{\hsize}{6.6cm}{\includegraphics[width=9.3cm]{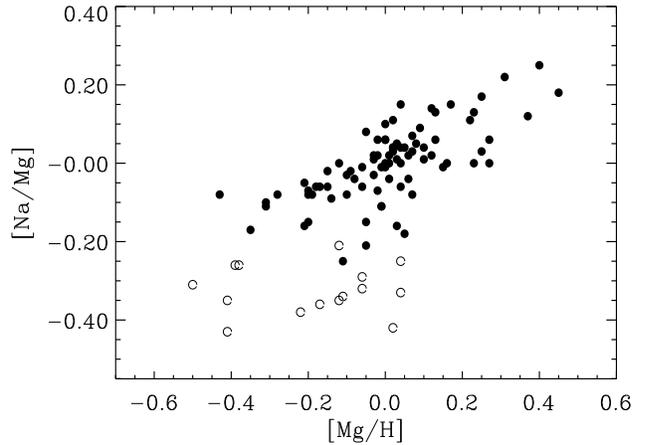}}
\caption[short title]{[Na/Mg] abundances ratios for NLTE analysis
as a function of [Mg/H]. Symbols are the same as in Fig. 2}
\end{figure}

Presently, the source of sodium remains to be confirmed. Thus it
is difficult to use sodium as a probe of Galactic chemical
evolution until the effects of individual stars can be quantified.
Theoretical sodium yields of SN II have not yet converged to
consistency among models of different authors (Woosley \& Weaver
\cite{WW95}; Umeda et al. \cite{UNN00}). Umeda et al.
(\cite{UNN00}) show the yields of sodium are much less for $Z=0$
compared to the $Z=0.02$ for a 20 $M_{\sun}$ model, while Woosley
\& Weaver (\cite{WW95}) predict that for a 22 $M_{\sun}$ SN model
the amount of sodium decreases with increasing metallicity in the
range of $Z=0$ to $Z_{\sun}$. Also, \emph{AGB stars are potential
\textbf{sites} for the production of primary sodium} (Mowlavi
\cite{M99}). In this scenario synthesis of sodium can occur in
hydrogen-burning shells during ON-cycling (Denissenkov \& Tout
\cite{DT00}). This process is supported by observations of red
giant stars in globular clusters that appear to be sodium-rich
(Kraft et al. \cite{KSS97}; Gratton et al. \cite{GBB01}).

Detailed modelling of the Galactic chemical evolution has been
attempted by many authors (e.g. Timmes et al. \cite{TWW95};
Goswami \& Prantzos \cite{GP00}). Based on Woosley \& Weaver's
(\cite{WW95}) metallicity-dependent yields, the first authors
calculate the behaviour of [Na/Fe] as a function of metallicity.
Their results predict that [Na/Fe] decreases from [Fe/H] $\sim$
$-3$ to $\sim -1.5$, while it increases from [Fe/H] $\sim$ $-1.5$
to $\sim -0.5$ (see their Fig. 17). However using the same yields
with an iron yield reduced by a factor of 2, Goswami \& Prantzos'
calculation explains the decline of [Na/Fe] from [Fe/H]$\sim -1$
to [Fe/H] $\sim -2$ reasonably well. Their prediction suggests an
ever-decreasing [Na/Fe] toward the lower metallicity regime. They
used a different initial mass function (IMF), and a different halo
model.

Our observational results provide some implications for the
nuclesynthesis of sodium. We note that in halo stars sodium is
significantly underabundant relative to iron, and [Na/Fe] (or
[Na/Mg]) increases with metallicity (Fig. 5). In the region of
overlapping metal abundances [Na/Fe] shows a clear distinction
between halo and thick-disk stars. The linear relationship
indicates that sodium production is connected to the metallicity
of the progenitor SN as suggested by Nissen \& Schuster
(\cite{NS97}). This result agrees with nucleosynthesis
calculations of massive stars by Umeda et al. (\cite{UNN00}), but
it is at variance with the calculations by Woosley \& Weaver
(\cite{WW95}). We note that there are no extremely metal-poor
stars in the sample of Gehren et al. (\cite{GLS04}), whereas
McWilliam et al. (\cite{MPS95}) find that, for stars with [Fe/H]
$\leq -3$, [Na/Fe] rises with decreasing metal abundance. Based on
this result Tsujimoto et al. (\cite{TSY02}) argued that sodium
yields decrease with increasing $Z$ for the very metal-poor stars.
However, nearly all the stars are giants and the abundances were
derived from the Na D (5890/5896) lines, which are very sensitive
to NLTE effects (see Sect. 5.1; Takeda et al. \cite{TZT03}). There
are two possibilities to explain the high [Na/Fe] values of
McWilliam et al.,
\begin{itemize}
\item
NLTE corrections for Na D lines are large for metal-poor cool giants, so if it is
considered, the [Na/Fe] ratios would show an abundance deficiency (however, see
Takeda et al. \cite{TZT03}).
\item
sodium is also produced in metal-poor giant stars (Goriely \& Siess \cite{GS01}),
which could explain why the Pb-stars also have a large sodium content (Van Eck
et al. \cite{VGJ01}).
\end{itemize}
A definitive confirmation of the latter assumption will require
more detailed nucleosynthesis calculations. It is highly desirable
to include the sample of ultra-low metallicity stars for
establishing the behaviour of [Na/Fe] in this region, which is
important for investigating the history of the early Galaxy as
well as for constructing the model of Galactic chemical evolution.
\begin{figure}
\resizebox{\hsize}{6.6cm}{\includegraphics[width=9.3cm]{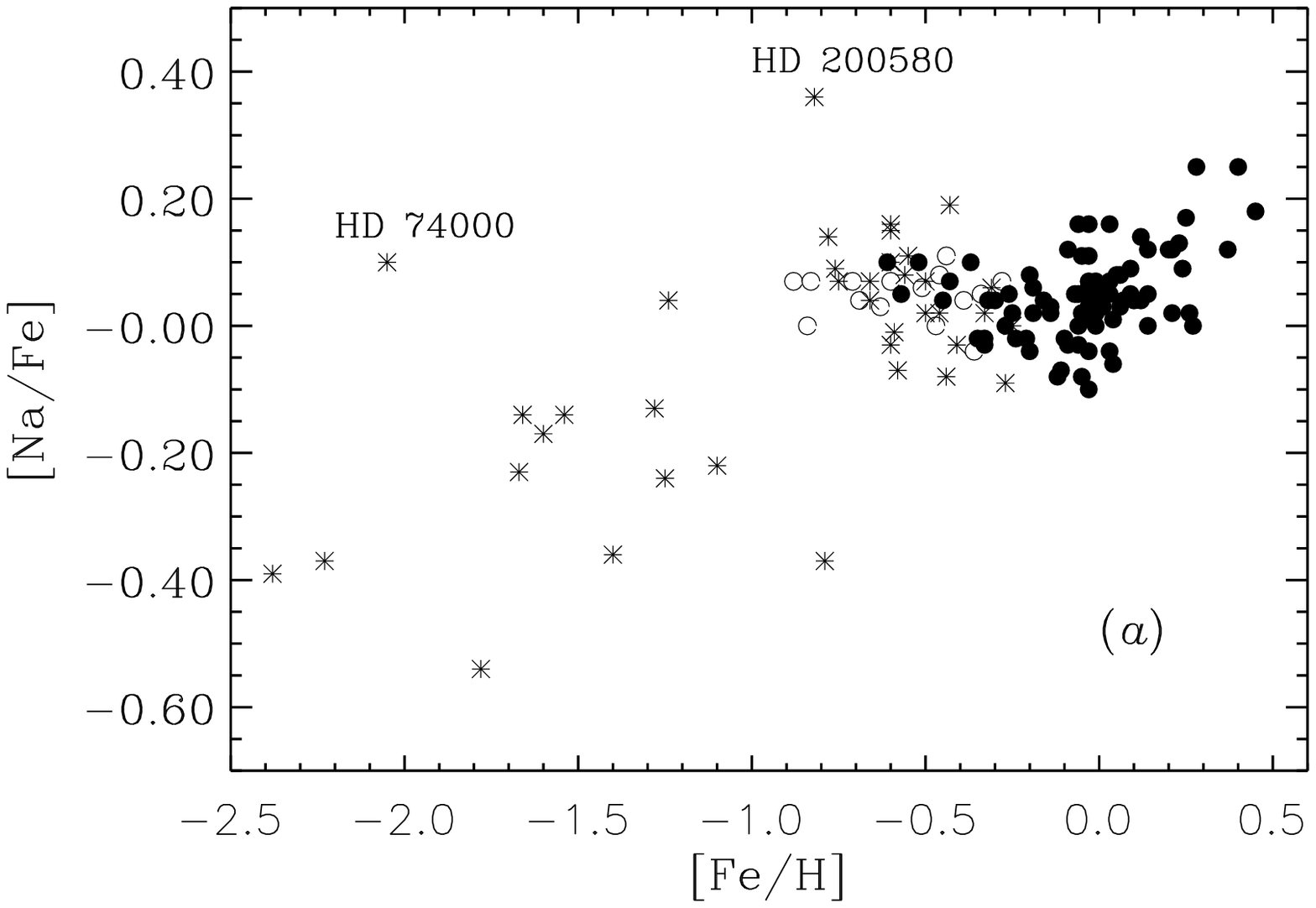}}
\resizebox{\hsize}{6.6cm}{\includegraphics[width=9.3cm]{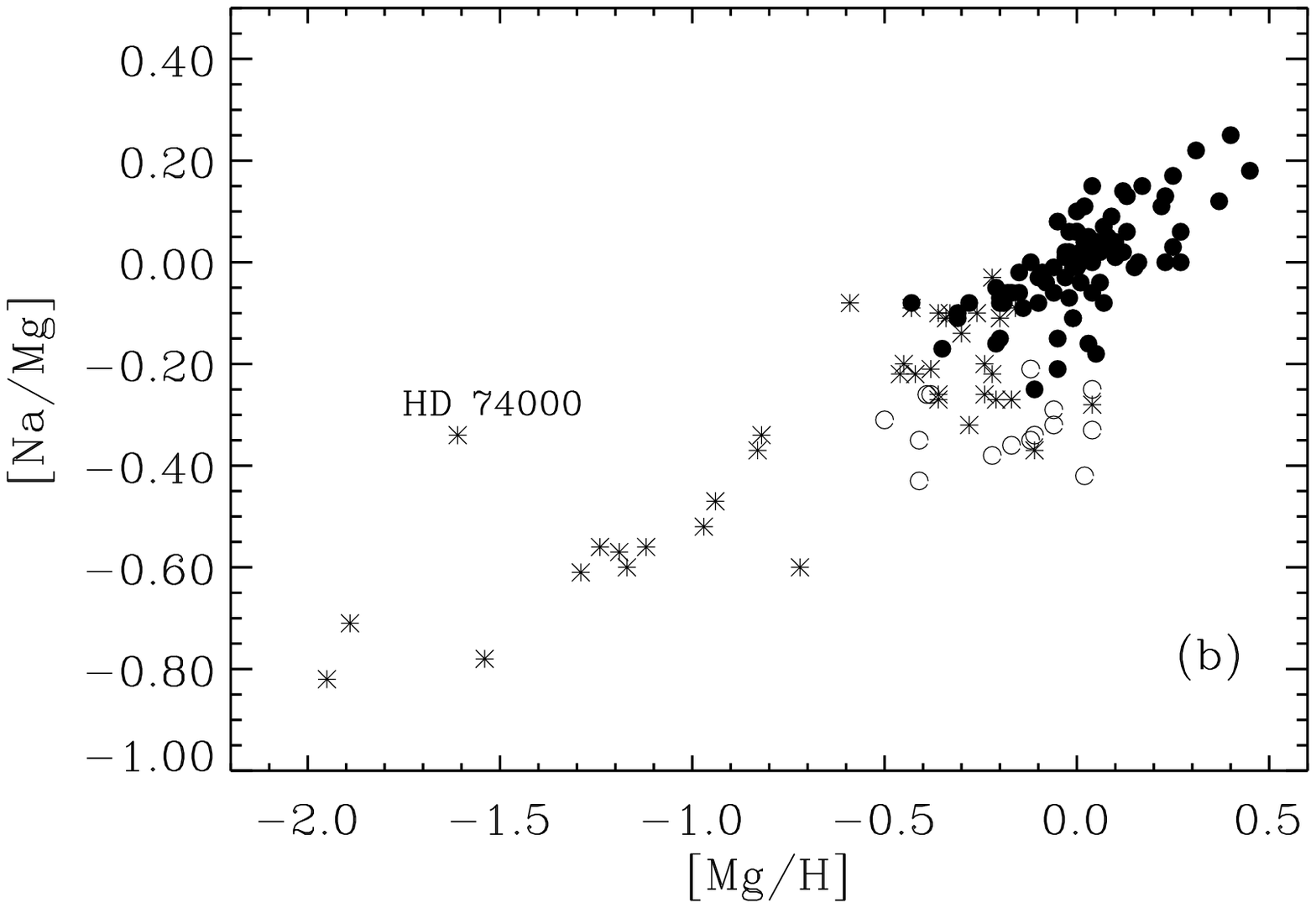}}
 \caption[short
title]{Top: abundance ratios [Na/Fe] for NLTE analysis as a
function of [Fe/H]. Bottom: abundance ratios [Na/Mg] for NLTE
analysis as a function of [Mg/H]. Filled and open circles refer to
the thin- and thick-disk stars in this paper, respectively.
Asterisks refer to the stars from Gehren et al. \cite{GLS04} }
\end{figure}

Based on [Eu/Ba] ratios, Mashonkina et al. (\cite{MGT03})
suggested that during the active phase of the thick disk formation
evolved low mass stars may enrich the interstellar gas. Our
observational results show that [Na/Fe] values are about solar,
possibly being slightly overabundant in the thick-disk stars,
which would confirm the suggestion that AGB stars may contribute
some additional sodium. However, metallicity-dependent SN yields
of sodium can not be excluded.

While the origin of the ``upturn" for the metal-rich stars is
still unclear, Feltzing \& Gustafsson (\cite{FG98}) argued that
metallicity-dependent SN yields may play a role. Though a rather
insufficient fraction of stars supersolar metal abundances
prevents us from making any definitive argument, we suggest that
quite large amounts of Na may be produced by proton capture on
$^{22}$Ne nuclei for AGB stars (Denissenkov \& Denissenkova
\cite{DD90}).

\subsection{Abundance ratios, kinematics, ages and the Galactic stellar
populations}

It is important to distinguish the individual membership of stars
in our sample. We are aware that such a discussion requires more
than the interpretation of abundance ratios alone. Following
Fuhrmann's (\cite{F98}, \cite{F00}) work we use [Mg/Fe] $=0.25$ as
a first step to discriminate between thin- and thick-disk stars.

\subsubsection{Abundance ratios and population membership}
Fuhrmann (\cite{F98}) attributed the scatter in [Mg/Fe] to a mixing of thin- with
thick-disk stars and a different chemical evolution of these two stellar populations
(see also Gratton et al. \cite{GCM00}; \cite{GCC03} for remarks on O and Mg with
respect to Fe). He suggested that the stars with higher [Mg/Fe] at a given [Fe/H] are
old thick-disk stars (see also Reddy et al. \cite{RTL03}).

We show the distribution of [Na/Fe] versus [Mg/Fe] for the two
populations of thick- and thin-disk stars in Fig. 6. Even
comparing abundance alone, Fig. 6 suggests differences among the
two populations: the thin-disk stars are well separated on the
left side.

\begin{figure}
\resizebox{\hsize}{6.6cm}{\includegraphics[width=9.3cm]{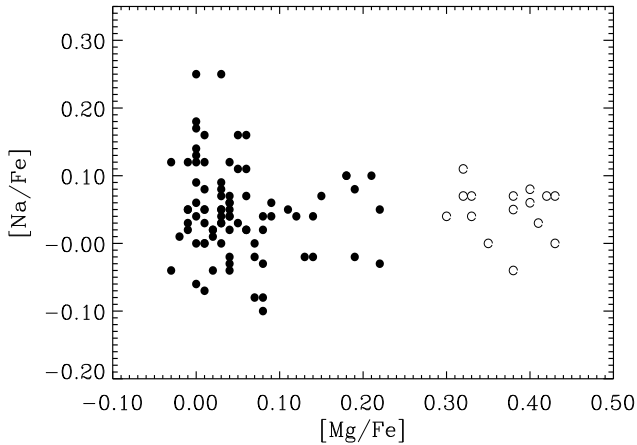}}
\caption[short title]{[Na/Fe] abundance ratios for NLTE analyses
as a function of [Mg/Fe]. The meaning of symbols is the same as in
Fig. 2} \vspace*{-0.3cm}
\end{figure}

\subsubsection{Kinematic properties and population membership}
The original classification of stellar populations was based much more on
kinematics than on anything else. It is therefore of high priority to check if
kinematic properties and abundance ratios have more in common than a coarse
change with the mean metal abundance itself. Fuhrmann (\cite{F00}; \cite{F02};
see also Ryan \& Smith \cite{RS03}) argues that in fact halo stars may
constitute an extremely small minority of all Galactic stars, whereas most of
the stars with intermediate kinematics instead belong to the thick-disk.
Fuhrmann (\cite{F00}) also suggested that there is some kinematical overlap of
the disk populations. Therefore the determination of Galactic kinematics
provides an important test.

Fuhrmann (\cite{F00}) has discussed in more detail the kinematical aspects of
our sample. The basic data are available in a number of catalogues which are
electronically accessible (e.g. via http://www.ari.uni-heidelberg.de/aricns).
The correction for the basic solar motion is that of Dehnen \& Binney
(\cite{DB98}), namely, $U_{0}/V_{0}/W_{0} = 10.00/5.25/7.17$ km s$^{-1}$.
Whenever these data are not provided, we have calculated the space velocities
following the model described by Chen et al. (\cite{CNZ00}).

The first inspection of Fig. 7 shows that the $V$ velocities of
thin-disk stars tend to be significantly larger than those of the
other populations, although there is some overlap between thin-
and thick-disk stars. The two most exceptional cases are the
thin-disk star HD\,52711 (V=$-98$ km s$^{-1}$) and the transition
star HD\,90508 (V=$-89$ km s$^{-1}$), and we will come back to
them later. The correlation between low [Na/Mg] ratios and low V
or high peculiar space velocities ($v_{spec}$) is also instructive
(Fig. 8). The kinematic status in the Toomre diagram of our sample
was shown by Fuhrmann (\cite{F00} in his Fig. 17).

We now come back to the transition stars HD\,90508 ($U/V/W
=31/-89/30$ km s$^{-1}$) and thin-disk object HD\,52711 ($U/V/W
=-17/-98/-13$ km s$^{-1})$: Fuhrmann (\cite{F00}) argues that
although HD\,90508 shows thick-disk kinematics, its chemistry and
age are more like thin-disk stars. HD\,52711 is thought to be a
thick-disk star based on its \emph{kinematics} (e.g. Ibukiyama \&
Arimoto \cite{IA02}), however, it is very young (6.8 Gyr) and
[Mg/Fe] is about solar (0.04), therefore it is also more like a
thin-disk star.

\begin{figure}
\resizebox{\hsize}{6.6cm}{\includegraphics[width=9.3cm]{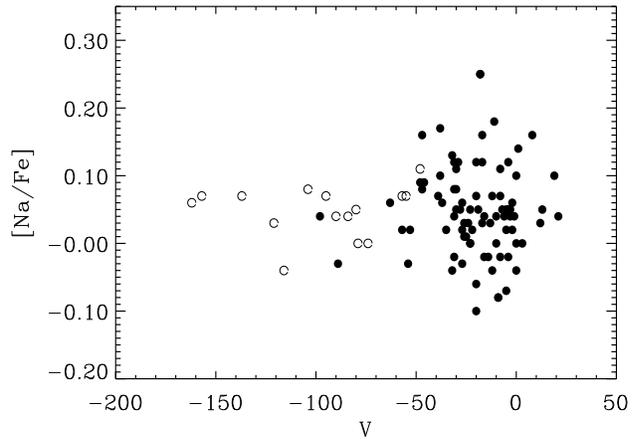}}
\resizebox{\hsize}{6.6cm}{\includegraphics[width=9.3cm]{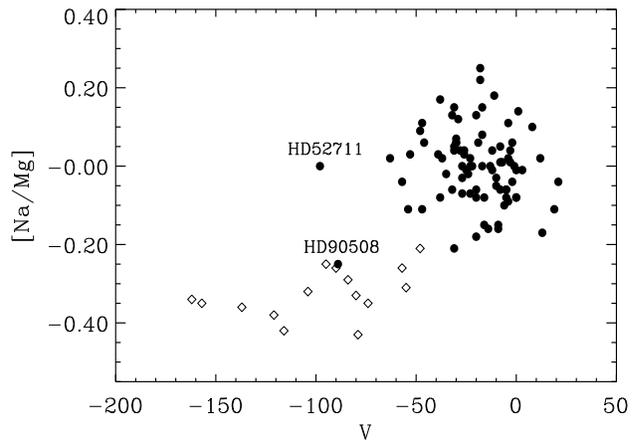}}
\caption[short title]{Correlation between Na abundance ratios and
orbital $V$ velocity components. Symbols are as in Fig. 2}
\end{figure}

\begin{figure}
\resizebox{\hsize}{6.6cm}{\includegraphics[width=9.3cm]{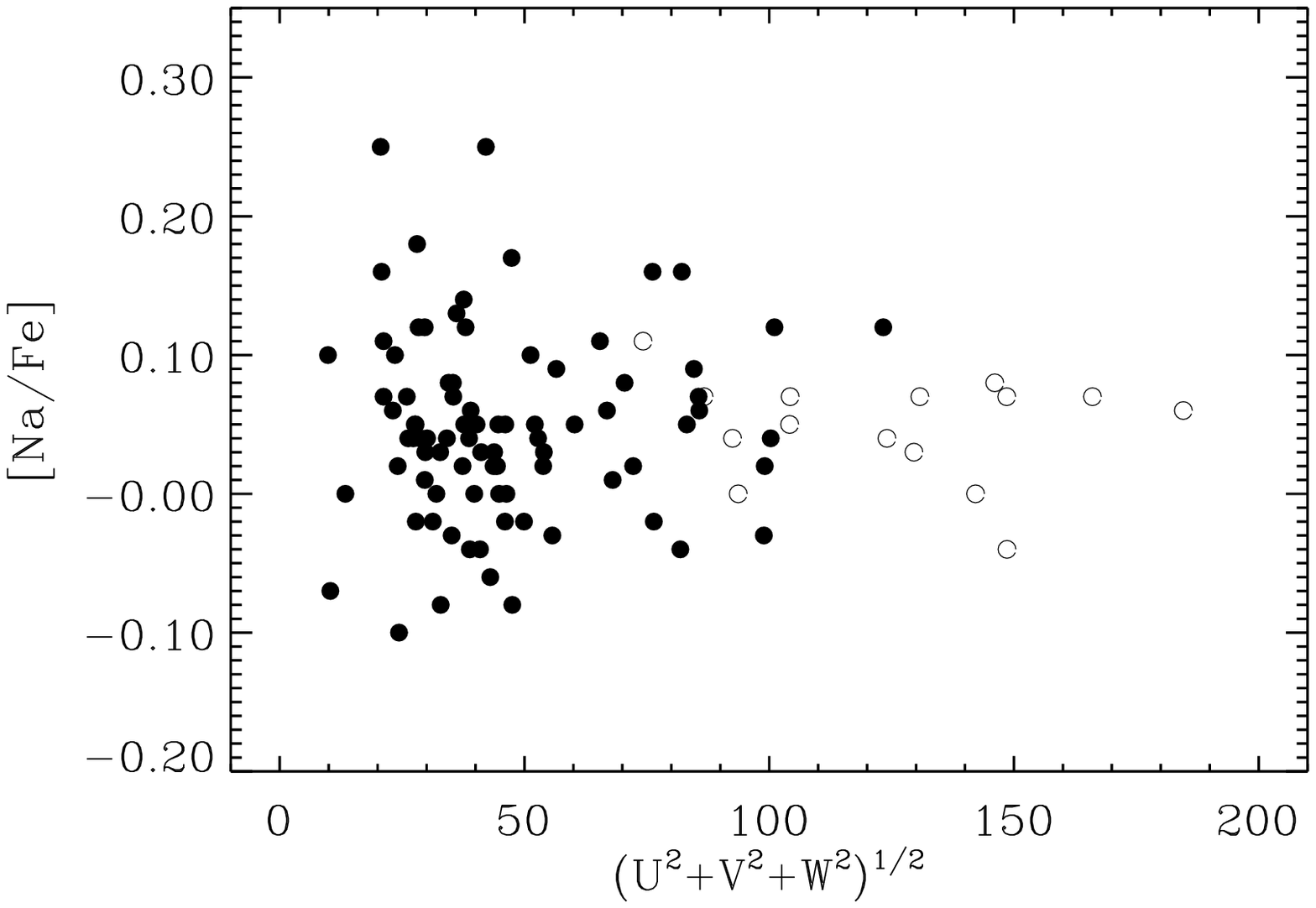}}
\resizebox{\hsize}{6.6cm}{\includegraphics[width=9.3cm]{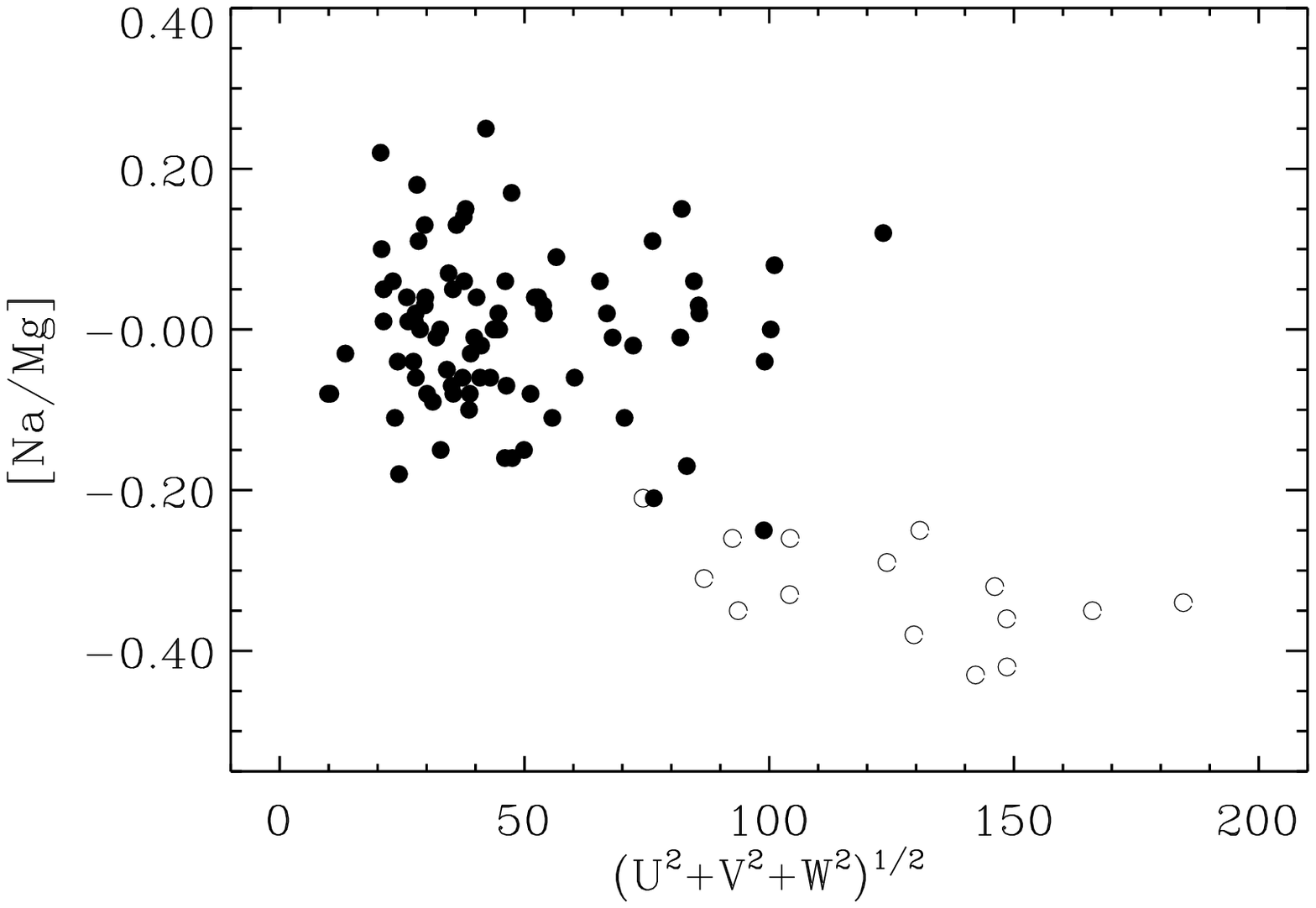}}
\caption[short title]{Correlation between [Na/Fe] abundance ratios
and peculiar space velocities $v_{spec} = (U^2+V^2+W^2)^{1/2}$
(top), and correlation between [Na/Mg] abundance ratios and
peculiar space velocities $v_{spec} = (U^2+V^2+W^2)^{1/2}$
(bottom). The meaning of the symbols is the same as in Fig. 2}
\end{figure}

\subsubsection{Stellar ages and population membership}
We now turn to the age aspects of analyzed stars. Gratton et al. (\cite{GCM00})
concluded that there is a gap in the distribution of [O/Fe] as expected from a
sudden decrease in star formation during the transition from the thick- to
thin-disk phase. Fuhrmann (\cite{F00}) suggested that \emph{the key feature to
identify thick- and thin-disk stars individually is provided by the stellar
ages, while there is a considerable age overlap between halo and thick-disk
stars}.

Using the stellar effective temperatures together with the
absolute magnitudes based on Hipparcos parallaxes, masses and
approximate ages can be interpolated according to [Fe/H] and
[$\alpha$/Fe] from adequate tracks of stellar evolution. Such
calculations have recently become available through the work of
VandenBerg et al. (\cite{VSR00}, \cite{VRM02}) and Yi et al.
(\cite{YKD03}). In this paper we adopt the Yonsei-Yale isochrones
(Yi et al. \cite{YKD03}) since the evolutionary tracks of
VandenBerg contain only metal-poor ([Fe/H] $\leq -0.31$) and low
mass stars (M$_{sun} \leq 1.20$). The Yonsei-Yale isochrones are
calculated with new OPAL opacities and Kurucz model atmospheres
for a set of metallicities $Z =$ 0.00001, 0.0001, 0.0004, 0.001,
0.004, 0.007, 0.01, 0.02, 0.04, 0.06, 0.08, and [$\alpha$/Fe] =
0.0, 0.3, 0.6. The mixing length and helium enrichment rate were
fixed to $\alpha \equiv l/H_{p}=1.7431$ and $\Delta$Y/$\Delta
Z=2.0$, respectively. The full set of stellar models and a FORTRAN
package that works for mass, metallicity and $\alpha$-enhancement
interpolation are available from the authors.

The ages derived here are based on the assumption that [O/Mg]$ = 0$. From Fig. 9 we can
see that all the thin-disk stars do not exceed $\sim 9$ Gyr, whereas with the exclusion
of HD\,165401 all thick-disk stars are older than $\sim 9$ Gyr. Fuhrmann (\cite{F98})
identified this star as a thick-disk member with an age of about 10 Gyr, but in his paper
II (Fuhrmann \cite{F00}) he noted that it is a chromospherically active, high-velocity
star, so it was excluded from his later analysis.

As already pointed out by VandenBerg et al. (\cite{VRM02}) and iterated by
Gehren et al. (\cite{GLS04}), a significantly higher [Mg/Fe] ratio could reduce
the ages of thick-disk and halo stars by as much as 3 Gyr. For comparison, we
also calculated the ages for these stars without $\alpha$-enhancement. Then our
result is fairly in agreement with Fuhrmann's, namely that all the thick-disk
(excluding HD\,165401) and halo stars are older than 12 Gyr, while all the
thin-disk stars are younger than 9.5 Gyr except the three transition stars
(HD\,90508, HD\,143761 and HD\,187923). Therefore Fuhrmann's result that
\emph{the key feature to identify thick- and thin-disk stars individually is
provided by the stellar ages} still holds. However, the 3 Gyr long star
formation gap between thin- and thick-disk is not confirmed here.

The three transition stars (HD\,90508, HD\,143761 and HD\,187923) have ages
about 8.8 Gyr, thus they are as old as the oldest thin-disk stars. They would be
about 10.5 Gyr old if no $\alpha$-enhancement were considered.

\begin{figure}
\resizebox{\hsize}{6.6cm}{\includegraphics[width=9.3cm]{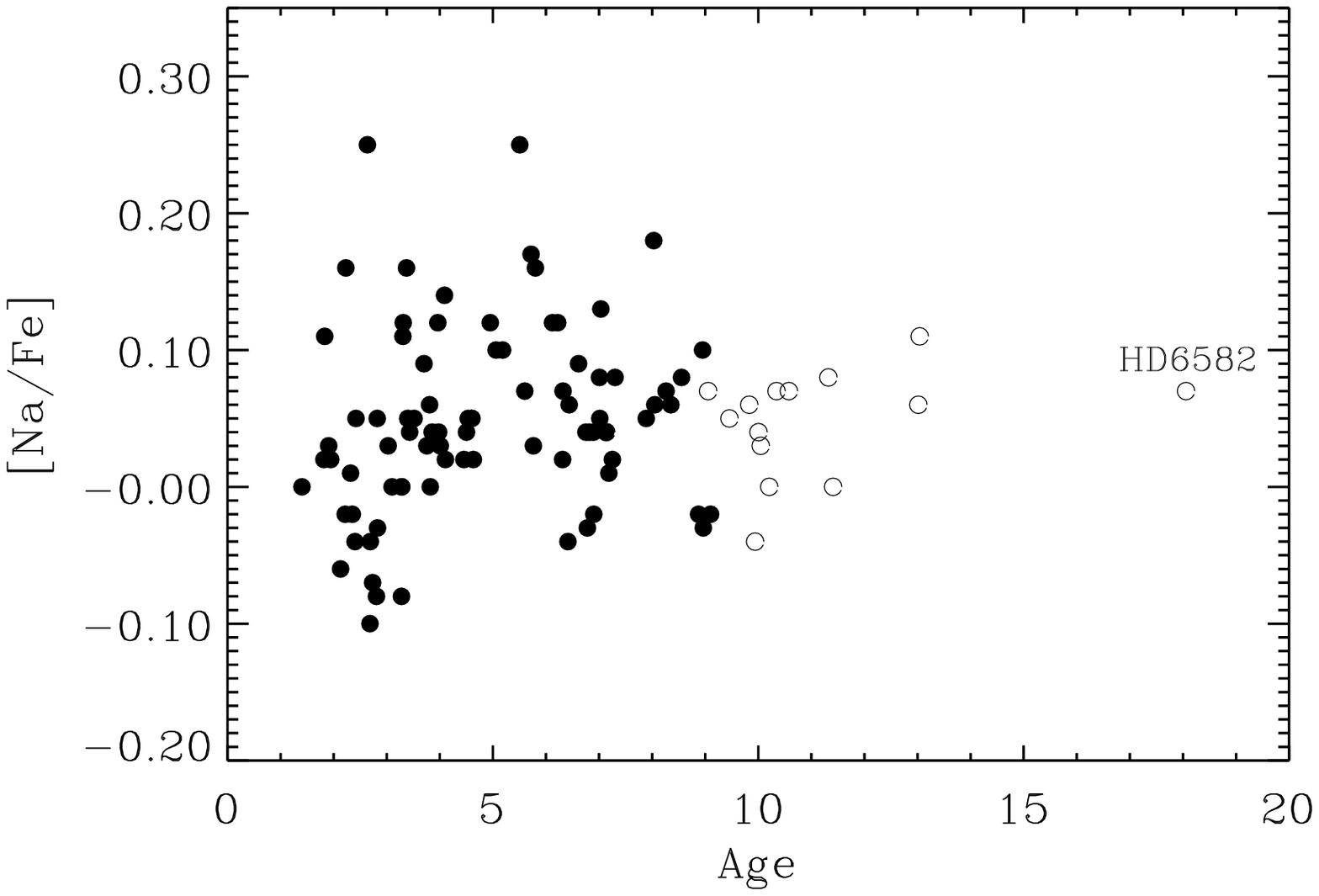}}
\resizebox{\hsize}{6.6cm}{\includegraphics[width=9.3cm]{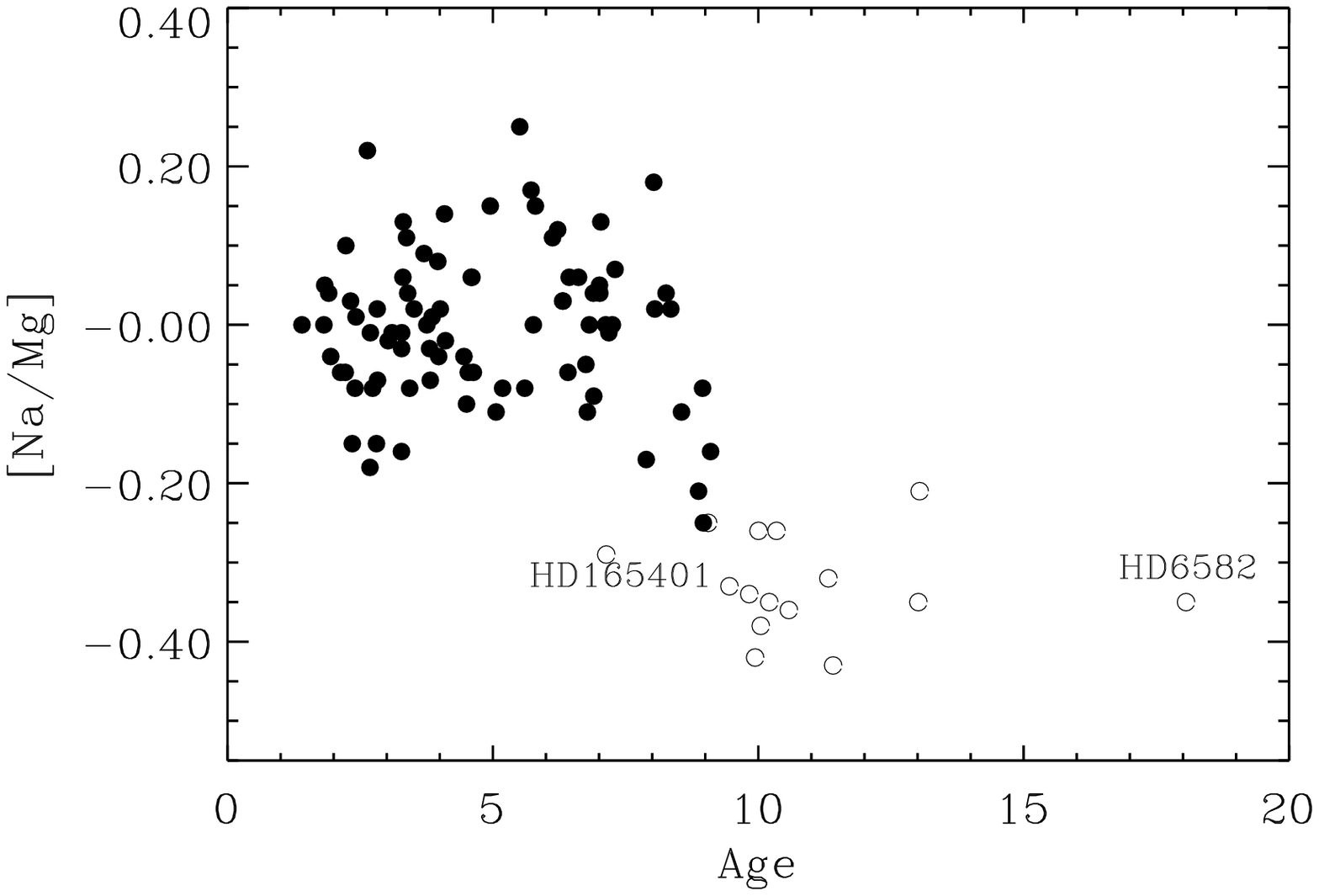}}
\caption[short title]{[Na/Fe] ratios vs. stellar ages (top), and
[Na/Mg] abundance ratios vs. stellar ages (bottom). Ages are in
Gyr. Symbols have the same meaning as in Fig. 2}
\end{figure}

We also note that the age of the thick-disk star HD\,6582 seems to be very high
($>$15 Gyr); Gehren et al. (\cite{GLS04}) have discussed this in detail.

\section{Conclusions}
We have determined sodium abundances for some 90 nearby stars, spanning the
range -0.9 $<$ [Fe/H] $<+$0.4. All abundances are derived from NLTE statistical
equilibrium calculations. Based on our results we come to the following
conclusions:
\begin{enumerate}
\item
The [Na/Fe] ratios are about solar for thin- and thick-disk stars.
The ``upturn" of [Na/Fe] for metal-rich stars is reproduced.
\item
The NLTE effects are different from line to line. The 6154/6160 lines are
comparatively less sensitive to NLTE effects, while the Na D and near-IR line
pairs show large NLTE effects. Large departures from LTE appear for warm
metal-poor subgiant stars.
\item
Our results suggest that the sodium yields increase with
metallicity, and quite large amounts of sodium may be produced by
AGB stars.
\item
Combined with the abundance ratios, kinematics and stellar ages, we identify
individual thick- and thin-disk stars. Fuhrmann's suggestion that stellar ages
provide a key feature in discriminating the two disk populations is supported.
\end{enumerate}

It would be important to perform NLTE abundance determinations for
both Mg and Al for these program stars to confirm our
identification, because both Al and Mg show large NLTE effects for
metal-poor stars (Gehren et al. \cite{GLS04}). It will be also
necessary to investigate some extremely metal-poor stars.

\begin{acknowledgements}
SJR thanks the Institute of Astronomy and Astrophysics of Munich University for
warm hospitality during a productive stay in 2002. The authors thank Klaus
Fuhrmann for his kind permission to use the reduced FOCES spectra of the stars
investigated in this paper. This research was supported by the "Deutsche
Forschungsgemeinschaft under contract Ge 490/26-1 and  by the National Natural
Science Foundation of China under grants No.10173014, No.10173028 and NKBRSF
1999075406.
\end{acknowledgements}

\end{document}